\documentclass[twocolumn,aps,prb,superscriptaddress]{revtex4}
\usepackage{graphicx}
\begin{document}

\title{Structure, reactivity and electronic properties of V-doped Co clusters}
\author{Soumendu Datta,$^1$ Mukul Kabir,$^2$ Tanusri Saha-Dasgupta,$^1$ and Abhijit Mookerjee$^1$}

\affiliation{$^1$Department of Material Sciences, S.N. Bose National Centre for Basic Sciences, JD Block, Sector-III, Salt Lake 
City, Kolkata 700 098, India \\ 
$^2$Department of Materials Science and Engineering, Massachusetts Institute of Technology, Cambridge, Massachusetts 02139, USA }

\date{\today}

\begin{abstract}
Structures, physical and chemical properties of V doped Co$_{13}$ clusters have been studied in detail using density functional
 theory based first-principles method. We have found anomalous variation in stability of the doped clusters with increasing V 
concentration, which has been nicely demonstrated in terms of energetics and electronic properties of the clusters. Our study
explains the nonmonotonic variation in reactivity of Co$_{13-m}$V$_m$ clusters towards H$_2$ molecules as reported
 experimentally [J. Phys. Chem. {\bf 94}, 2744 (1990)]. Moreover, it provides useful insight into the cluster geometry and chemically
 active sites on the cluster surface, which can help to design better catalytic processes.
\end{abstract}
\maketitle

\section{\label{sec:intro}Introduction}

The interest in the studies of clusters is largely because of  their possible technological applications 
which include the possibilities of developing novel cluster-based materials using the size dependence of
their properties. Doping of clusters is an important possibility in this direction.
 In recent times the fabrication of  alloy clusters of different sizes with well defined, controlled
 properties by varying the composition and atomic ordering, has caught considerable attention.\cite{review1, review2}
 Bimetallic alloy clusters have been known and exploited for last few years in various applications, specially in the 
catalytic reactions.\cite{ref1, toshima} Varying the ratio of the two constituents, one can alter the surface structures, compositions
 and segregation properties.\cite{ruban, bozzolo} In this way, it is possible to tune the chemical reactivity at the surface of an
alloyed cluster.\cite{ref2, molenbroek} Considering the huge possibility of using nano clusters in catalysis, the study of cluster reactivity
 has become an interdisciplinary topic of present day research.\cite{new1, new2, new3, new4}

 About two decades ago, Nonose {\it et al} measured the reactivity towards H$_2$ of
 bimetallic Co$_n$V$_m$ ($n$ $>$ $m$) clusters using laser vaporization technique and reported strong cluster size and composition
dependence.\cite{nonose} Both V and Co are 3$d$ transition metals. The substitution of Co by V atoms, one by one, should
 increase the reactivity of the alloy cluster towards H$_2$ molecules, as V, an early transition metal, has a high
 reactivity towards H$_2$ in contrast to Co which has relatively low reactivity.\cite{ref3}
Interestingly, it was found that the reactivity increased gradually, as expected, with the substitution of Co atom
 by V atom one by one in the Co$_n$ clusters ($n$ $<$ 13), but for the Co$_{13}$, there was a remarkable decrease in reactivity when
 a single Co atom was substituted by a V atom. However, the reactivity increased
 as the number of exchange V-atoms increased further up to  $m =$ 3, while the fourth V-atom
 substitution did not increase the reactivity any more. 
In view of the high reactivity of  elemental V, this sudden
drop in reactivity of the Co$_{12}$V cluster was rather surprising. The authors speculated a plausible icosahedral
 structure for the Co$_{12}$V cluster with the active V atom at the cage center. Therefore, the V atom, being shielded
geometrically from $H_2$ by the twelve surface Co atoms, might have less chance to interact with H-atoms. The chemisorption
 reactivity of cationic Co$_{13-m}$V$_m^+$ clusters \cite{cationicCoV} and anionic
 Co$_{13-m}$V$_m^-$ clusters \cite{anionicCoV} also shows similar type of variation as that of neutral Co$_{13-m}$V$_m$
 clusters, which hints towards the dominant effect of geometric structure as compared to the electronic structure. For clusters,
the ionization potential (IP) depends on the position of the highest occupied molecular orbital (HOMO) level. For the pure metal clusters, Whetten {\it et al} have postulated that the reaction rate for cluster-H$_2$ dissociative chemisorption is determined by
 the charge transfer from the HOMO to the lowest unoccupied molecular orbital (LUMO) of the reactant gas H$_2$, and an anti-correlation
 between IP and reaction coefficient could be observed.\cite{whetten} That means lower value of IP corresponds to higher reactivity and
vice versa. However, the ionization energies of the Co$_n$V$_m$ clusters, measured by Hoshino {\it et al} using
 photo-ionization spectroscopy show no such anti-correlation,\cite{hoshino} again demonstrating the importance of geometrical structure. A
rigorous first-principles study in terms of geometric and electronic effects is therefore very much needed to
understand this anomalous nature of reactivity of Co$_{13-m}$V$_m$ ($m =$ 0-4) clusters. In this report, we have carried
 out an {\it ab-initio} theoretical study on V doped Co$_{13}$ clusters. The whole study can be divided into three major parts :
first part consists of an exhaustive search for the minimum energy structures (MES) for cluster of each composition, followed
 by stability analysis of these MES in terms of various physical quantities in the second part,
 while the final part includes investigation of chemisorbed structures and understanding of cluster reactivity.

\section{\label{sec:methodology} Computational Details}

The calculations have been performed using density functional theory (DFT), within the pseudopotential
plane wave method as implemented in VASP code.\cite{kresse2} We have used the projector augmented wave (PAW) pseudopotentials \cite{blochl,kresse} and the Perdew-Bruke-Ernzerhof (PBE) exchange-correlation functional \cite{perdew} for spin-polarized generalized gradient approximation (GGA).
The 3$d$ and 4$s$ electrons were treated as the valence electrons for the transition metal elements and the wave
functions were expanded in the plane
wave basis set with the kinetic energy cut-off of 335 eV. The convergence of the cluster energies with respect to the cut-off value has
 been checked. Reciprocal
space integrations were carried out at the $\Gamma$ point. Symmetry
unrestricted geometry and spin optimizations were performed using the 
conjugate gradient and the quasi-Newtonian methods until all the force
components were less than a threshold value of 0.005 eV/{\AA}. In order to ensure the structural trends found in the optimized 
structures, we have also carried out Bohn-Openheimer molecular dynamics \cite{new5} within LDA for few specific clusters. Simple
cubic super-cells were used for cluster calculations with the periodic boundary conditions,
where two neighboring clusters were kept separated by at least 12 {\AA}
vacuum space to make the interaction between the cluster images
negligible.
The cohesive energy ($E_c$) of a Co$_n$V$_m$ alloy cluster is calculated with respect to the free atoms as
\begin{equation}
E_c({\rm {Co}}_n{\rm V}_m) = mE({\rm V}) + nE({\rm Co}) - E({\rm Co}_n{\rm V}_m)
\end{equation}
where $E$(Co$_n$V$_m$), $E$(Co) and $E$(V) are the total energies of Co$_n$V$_m$ cluster, an isolated Co atom
 and an isolated V atom respectively. One can also define the cohesive energy with respect to the Co and V bulk phases at equilibrium
 instead of isolated atoms. However, the cohesive energy variation with increasing V-concentration follows the same trend irrespective
 of the way of cohesive energy definition. The second difference in energy for fixed size ($n+m =$ constant) and
 variable composition is defined as
\begin{eqnarray}
\Delta_2E({\rm Co}_n{\rm V}_m) &=& E({\rm Co}_{n+1}{\rm V}_{m-1}) + E({\rm Co}_{n-1}{\rm V}_{m+1})\nonumber \\
 &-& 2E({\rm Co}_n{\rm V}_m)
\label{d2e}
\end{eqnarray}
It gives the relative stability of alloy clusters having nearby compositions.

\section{Structure}
Since chemical reaction with clusters takes place on its surface, the atomic arrangement and the 
composition of cluster surface play an important role in chemical reactivity.
Therefore, the first step towards the theoretical modeling of clusters is to determine their ground state
structure. In our earlier work,\cite{pureco} we studied the structure and magnetism of the pure Co clusters in detail. 
To obtain the MES having the optimized geometry as well as the optimized magnetic moment, we considered
 several probable starting geometries having closed packed atomic arrangement and allowed each of the geometries to relax for {\it all possible} collinear spin configurations of the atoms. We follow the same way of structural
 optimization here for the V doped Co clusters ( Co$_{13-m}$V$_m$; $m =$ 1-4). However, the situation for alloy clusters 
is quite cumbersome as one has to deal with a large number of starting geometries because of the presence of {\it homotops}
 (the term was first coined by Jellinek \cite{jellinek1,jellinek2}). The {\it homotops} have the same number of atoms, 
composition and geometrical structure, but differ in the arrangement of the doped atoms. As for example, a single geometrical
 structure of an A$_n$B$_m$ alloy cluster with fixed number of atoms ($N = m + n$) and composition ($m/n$ ratio),
 will give, in principle, $\frac{N!}{m! n!}$ {\it homotops}. Some of them, however, may be symmetry equivalent and the
 number of the inequivalent {\it homotops} will be somewhat less than the above mentioned value. So the variety of
 structures in alloy clusters is much richer than that of the pure clusters and the
potential energy surface of even a small cluster of few tens of atoms is of enormous complexity.

\begin{figure}[h]
\begin{center}
\rotatebox{0}{\includegraphics[height=2.5cm,keepaspectratio]{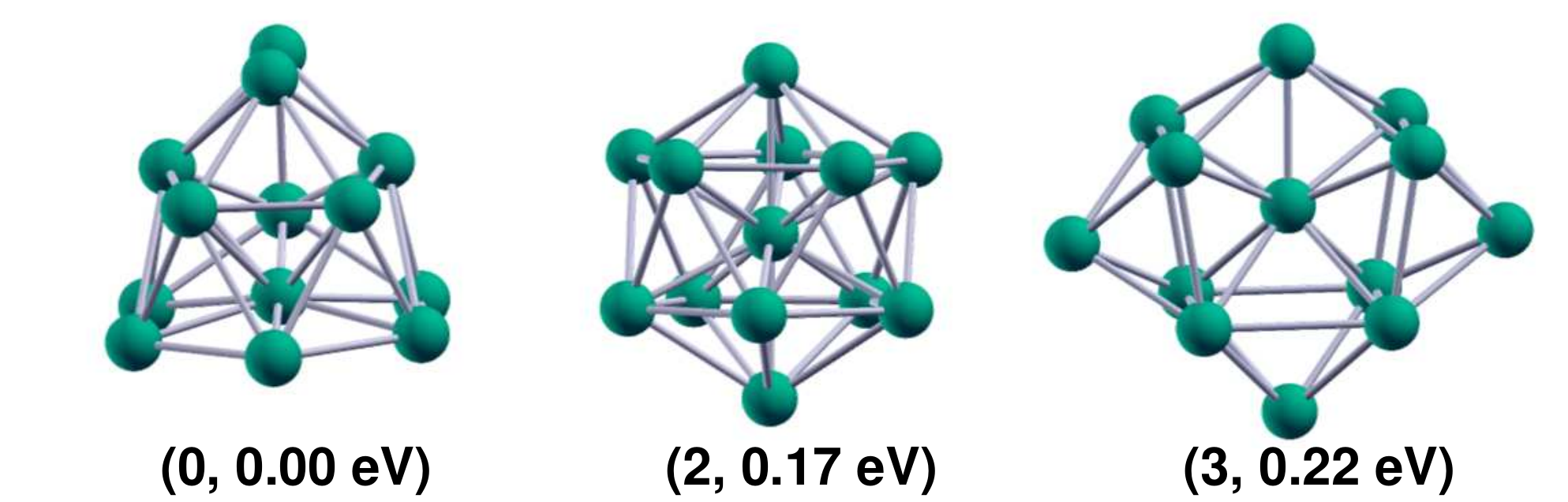}}
\caption{(Color online) Structures of optimal HCP, icosahedron and cub-octahedron of pure Co$_{13}$ cluster
(from left to right respectively). The first entry in the parenthesis gives the isomeric position (``0" means
 minimum energy state) and the second entry corresponds to relative energy to the minimum energy state.  }
\label{fig:strCo13}
\end{center}
\end{figure}

 It is generally found that the transition metal clusters prefer compact geometries to maximize the interaction between the rather localized
$d$ orbitals.\cite{alonso} There are three most compact and highly coordinated structures for 13 atoms cluster: icosahedron,
 cub-octahedron and hexagonal closed packed (HCP) geometries. 
We have therefore considered these three geometries as the most probable starting structures. For the pure Co$_{13}$
cluster, we found that the MES is a distorted HCP geometry with total magnetic moment of
 25 $\mu_B$ and total cohesive energy of 42.63 eV as mentioned in our previous work.\cite{pureco} The structure has 22
triangular faces and 33 edges (cf. Fig. \ref{fig:strCo13}). This hexagonal growth of the pure Co clusters is quite unusual compared
 to the trend seen in the clusters of its 3$d$ neighbors like Mn, Fe, Ni, which preferably adapt an icosahedral pattern.\cite{mn, fe, ni} 
Further study is needed to explore the unusual structural pattern of small pure Co clusters.\cite{future,thesis} Another distorted structure 
of HCP motif with total magnetic moment of 27 $\mu_B$ and 0.14 eV above the minimum energy state, is the first isomer. The optimal
 icosahedral structure of total spin 31 $\mu_B$ (structure having 20 triangular faces and 30 edges) is 0.17 eV higher than the
minimum energy state and emerges as the second isomer. The third isomer is a distorted cub-octahedron
 with total magnetic moment 25 $\mu_B$ and it is 0.22 eV above the minimum energy state.
 The structures of the ground state, second and third isomers are shown in Fig. \ref{fig:strCo13}. It is to be noted that for the pure
Co$_{13}$ cluster, all the atomic moments are ferromagnetically coupled and our predicted magnetic moment of the ground state structure
 is in good agreement with the experimental value which is 2.0 $\mu_B$/atom.\cite{co13_expt}

\begin{figure}[h]
\begin{center}
\rotatebox{0}{\includegraphics[height=2.1cm,keepaspectratio]{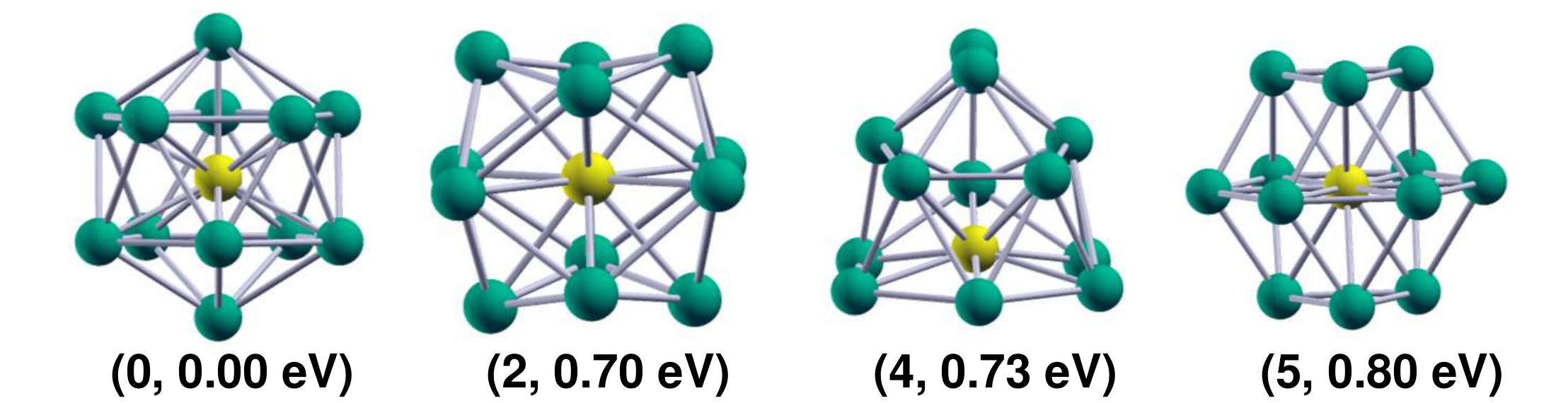}}
\caption{(Color online) Structures of optimal icosahedron, optimal cub-octahedron and two HCP geometries
 (from left to right) for Co$_{12}$V cluster. Green dot represents
 Co atom, while yellow dot represents V-atom. The entries in the parenthesis have same meaning as in Fig. \ref{fig:strCo13}.}
\label{fig:strCo12V}
\end{center}
\end{figure}

In the case of single V atom doped Co$_{12}$V clusters, we have again considered the starting geometries
of icosahedral, HCP and cub-octahedral symmetries. However, depending upon the position of the singly doped V-atom
  either at the center or on the surface, it can give rise to several {\it homotops}. Again, the surface atoms of a 
13-atoms cluster are not equivalent in terms of number of neighbors and their relative positions in case of HCP and 
cuboctahedral structures, while for icosahedral structure, all the surface atoms are equivalent. This consideration will
 give few more starting geometries. So we have considered all these possible geometrical
  structures and for each of them,  we have considered the {\it all possible} collinear spin alignments during relaxation.
 Upon optimization of geometry and spin degrees of freedom, we find that an icosahedral structure of
total magnetic moment 25 $\mu_B$, with V atom doped at the center position, is the MES. Now there are several interesting points
 to note in the context of the Co$_{12}$V cluster. First of all, the cohesive energy
 of the MES is considerably higher (by 1.29 eV) than that of the pure Co$_{13}$ clusters,
 indicating its exceptional stability. Secondly, the V doped Co-clusters are found to prefer the icosahedral growth pattern,
 instead of HCP growth pattern, as we observed in case of the pure Co-clusters. Therefore, just the single V-atom doping in
 Co$_{13}$ changes the equilibrium structure from HCP to icosahedron. This HCP to icosahedron structural transition
 has also been reported previously for Mn doping in Co$_{13}$ cluster.\cite{comn} Finally, the single V-atom likes
 to fit at the center of the icosahedron and it is ferrimagnetically coupled with the surface Co-atoms. The first isomer 
is also a central V-atom doped icosahedral Co$_{12}$V
 cluster, with total magnetic moment 23 $\mu_B$  and it is about 0.46 eV above the MES.
 Center V doped cub-octahedral Co$_{12}$V cluster with 23 $\mu_B$ magnetic
 moment which has energy 0.70 eV higher than the minimum energy icosahedral structure, is the second isomer. The third
isomer is a center V doped icosahedron with total magnetic moment of 27 $\mu_B$. Center V doping in case
of HCP cluster is less favorable and appears as the fifth isomer in our calculation with total magnetic
 moment 25 $\mu_B$ and energy 0.80 eV above the MES. We have also considered the optimal structure
 of Co$_{13}$ (distorted HCP) as the starting structure and substituted the most coordinated central Co-atom
 by a V-atom. After relaxation
 considering {\it all possible} spin alignments, it is found that the shape of the optimal structure remains
 more or less same as that of the optimal HCP Co$_{13}$ cluster and energetically, it is the fourth isomer with
total magnetic moment 23 $\mu_B$. The most probable surface V-doping has a cub-octahedral structure with
 total magnetic moment 21 $\mu_B$, but it is about 1.32 eV above the MES. The structure
 of ground state and few isomers for the Co$_{12}$V cluster are shown in Fig. \ref{fig:strCo12V}.

Because of the icosahedral growth preference of the Co$_{12}$V cluster and the presence of the singly doped V-atom at the central site,
 we consider only the icosahedral symmetry based geometries as the starting guess for more than one V atom doped clusters, in which
 one V atom is always at the center position, while the residual V atoms reside on the surface. For Co$_{11}$V$_2$ cluster, the second V atom
 can replace any of the surface Co atoms, as all the surface sites of a 13-atom icosahedron, are equivalent. After
 relaxation for {\it all possible} spin configurations, we find that the MES has total
 magnetic moment of 19 $\mu_B$ and total cohesive energy of 43.22 eV. The first and second isomers have magnetic moments of
 21 $\mu_B$ and 17 $\mu_B$ and they are 0.10 eV and 0.25 eV above the minimum energy state respectively. In the MES,
 the center V-atom is again ferrimagnetically coupled with the surface Co atoms and has much lower local magnetic moment as it was in the case of the Co$_{12}$V cluster, while the surface V atom has the maximum local magnetic moment and it is also ferrimagnetically coupled with the other surface Co atoms.

\begin{figure}[h]
\begin{center}
\rotatebox{0}{\includegraphics[height=2.9cm,keepaspectratio]{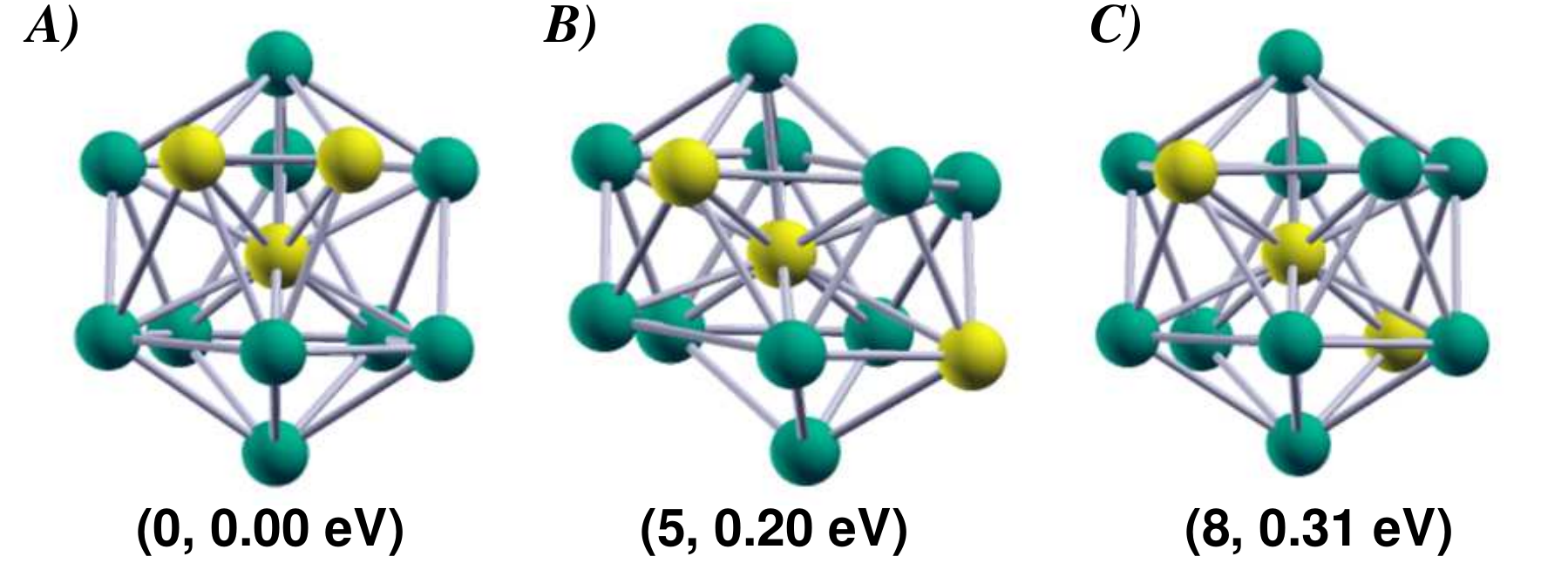}}
\caption{(Color online) $A$, $B$, $C$ represent the three inequivalent {\it homotops} in icosahedral Co$_{10}$V$_3$ structure having
 one V-atom always at the center. The three structures are the optimal structures for the respective three types.
 The entries in the parenthesis have same meaning as in Fig. \ref{fig:strCo13}. Color convention for atoms is same as in Fig. \ref{fig:strCo12V}. }
\label{strCo10V3}
\end{center}
\end{figure}

For Co$_{10}$V$_3$ cluster, depending upon the different positions of the two surface V atoms, there are
 three possible icosahedral structures as shown in Fig. \ref{strCo10V3}. However, the structure of type $A$ where the two surface V atoms
 are closest to each other, appears as the most favorable one. This type $A$ icosahedral structure with total magnetic
moment of 21 $\mu_B$ is the minimum energy state with ferrimagnetic alignment of the central V-atom and ferromagnetic alignment
 of the surface V-atoms with the surface Co-atoms. Both the central and the surface V-atoms have small values of local magnetic moments in this case.
 There are several closely spaced isomers of type $A$ icosahedron with magnetic moments of 19 $\mu_B$, 13 $\mu_B$, 23 $\mu_B$ and 
17 $\mu_B$ which are just 0.04 eV, 0.08 eV, 0.09 eV and 0.15 eV above the MES. We find that the optimal structures
 of type $B$ and type $C$ are 0.20 eV and 0.31 eV above the MES, respectively and
both of them have same total magnetic moment of 13 $\mu_B$.

\begin{figure}[h]
\begin{center}
\rotatebox{0}{\includegraphics[height=2.6cm,keepaspectratio]{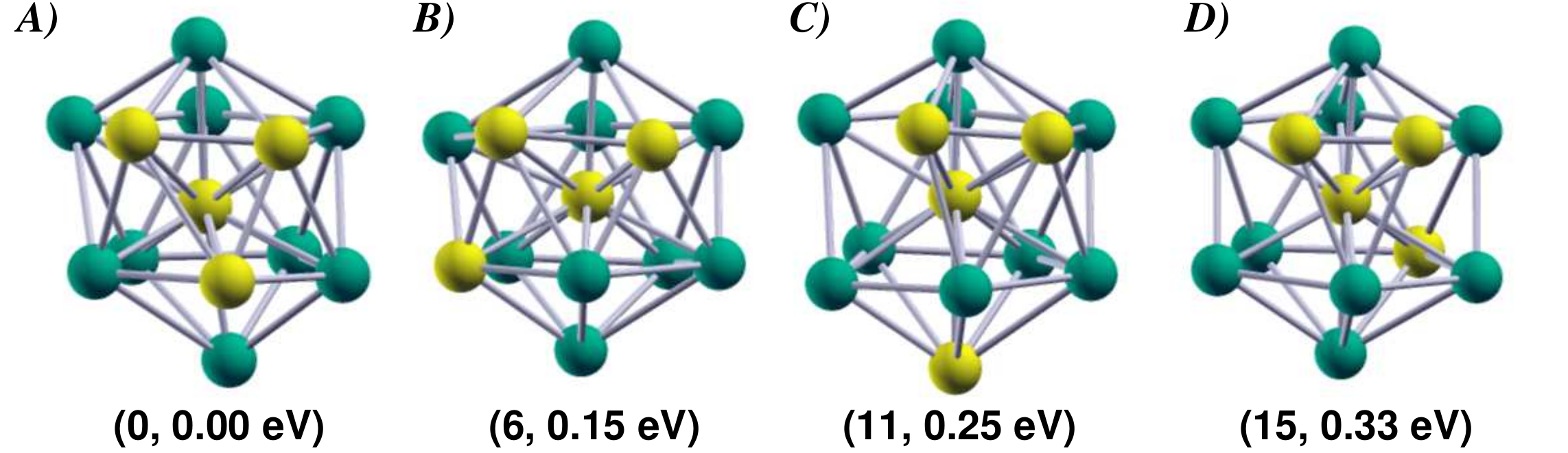}}
\caption{(Color online) $A$, $B$, $C$, $D$ represent the four probable {\it homotops} in icosahedral Co$_{9}$V$_4$ structure having
 one V-atom at the center and two of the remaining three V-atoms are closest to each other on the surface, while the
 position of the fourth V-atom is varied. The four structures represented above are the optimal structures for
 the respective four types. The entries in the parenthesis have same meaning as in Fig. \ref{fig:strCo13}. Color convention for atoms is same as in Fig. \ref{fig:strCo12V}.}
\label{strCo9V4}
\end{center}
\end{figure}

For Co$_{9}$V$_4$, there are three V atoms on the surface. Considering two surface V atoms at closest
 to each other (following the ground state configuration of Co$_{10}$V$_3$), different positions
 of the third surface V atom can give rise to the four icosahedral configurations as shown in Fig. \ref{strCo9V4}.
After optimization, it is found that the type $A$ is the most favorable structure where all the three surface V atoms are closest to each 
other and form an octahedron with the central V atom. The optimal structure of type $A$ icosahedron
 has total magnetic moment of 15 $\mu_B$ and  total cohesive energy of 42.35 eV. Another
type $A$ icosahedrons with magnetic moments of 13 $\mu_B$, 19 $\mu_B$, 21 $\mu_B$ and 17 $\mu_B$ are
just 0.03 eV, 0.07 eV, 0.09 eV and 0.13 eV above the MES. The optimal type $B$, type $C$ and type $D$ icosahedrons
have total magnetic moments of 15 $\mu_B$, 13 $\mu_B$ and 15 $\mu_B$ respectively and they are 0.15 eV,
 0.25 eV and 0.33 eV above the MES.

 The results of the structural optimizations, therefore, can be summarized as, unlike an HCP growth pattern of the pure Co cluster, the V doped
 Co$_{13}$ clusters prefer to adopt an icosahedral packing. In such clusters, the most coordinated central
 site is occupied by a V-atom, while the residual V-atoms sit on the surface. The surface V-atoms like to
 be closer to each other to form a group, thereby imparting more distortion to the structure and significantly alter the
 local surface charge density. The central V-site in the MES of all compositions, is
 always ferrimagnetically coupled and has lower magnetic moment. On the other hand, the surface V-atoms can be ferrimagnetically or
 ferromagnetically coupled with the surface Co-atoms and their local magnetic moments can be as low as the central V-site or as high as
 the surface Co-site, depending upon the distribution of surface charge density in the clusters having different amount of V-doping. 
Our prediction of icosahedral geometry for the minimum energy states of V doped Co clusters is in accordance with the speculation 
of a closed shell geometry around the central site in Ref. 13 and Ref. 15.\\
To have more confidence about the structural trend as observed above, we have carried out few {\it ab-initio}
 molecular dynamics (MD) study for the Co$_{12}$V and Co$_{10}$V$_3$ clusters. To determine the lowest energy structures, we have
 heated up the clusters to 2000 K (which is close to the melting temperature 1768 K for bulk Co and 2183 K for bulk V).
 The clusters have then been maintained at this temperature for at least 6 ps and then allowed to cool again to 0 K. 
The cooling process has been done for 24 ps and maintained a very slow cooling rate of 1 K per one iteration throughout 
the process. The main results obtained from the MD run are : (a) it prefers to adopt the icosahedral pattern with same
 magnetic alignment of atoms as we predicted from zero temperature relaxation, (b) For the Co$_{12}$V cluster, the V atom
 occupies the central position, (c) the surface V atoms for the Co$_{10}$V$_3$ cluster, prefer to be closest to each other. These results
 provide reassurance and more confidence in our zero temperature relaxation.

\section{Stability analysis}
 The observed atomic arrangement of the MES of the Co$_{13-m}$V$_m$ clusters as described
 in the previous section depends critically on the balance of several parameters like the relative
strengths of various kinds of bonds present in a cluster structure, the relative atomic sizes, the amount of charge transfer between
 two different species of atoms, the energy gap between the HOMO and the LUMO, abundance of states near Fermi energy etc. Below, we try to understand
 the structural stability of the clusters in terms of these parameters.

\subsection{Cohesive energy}
\begin{figure}[t]
\begin{center}
\vskip 0.73cm
\rotatebox{0}{\includegraphics[height=6.2cm,keepaspectratio]{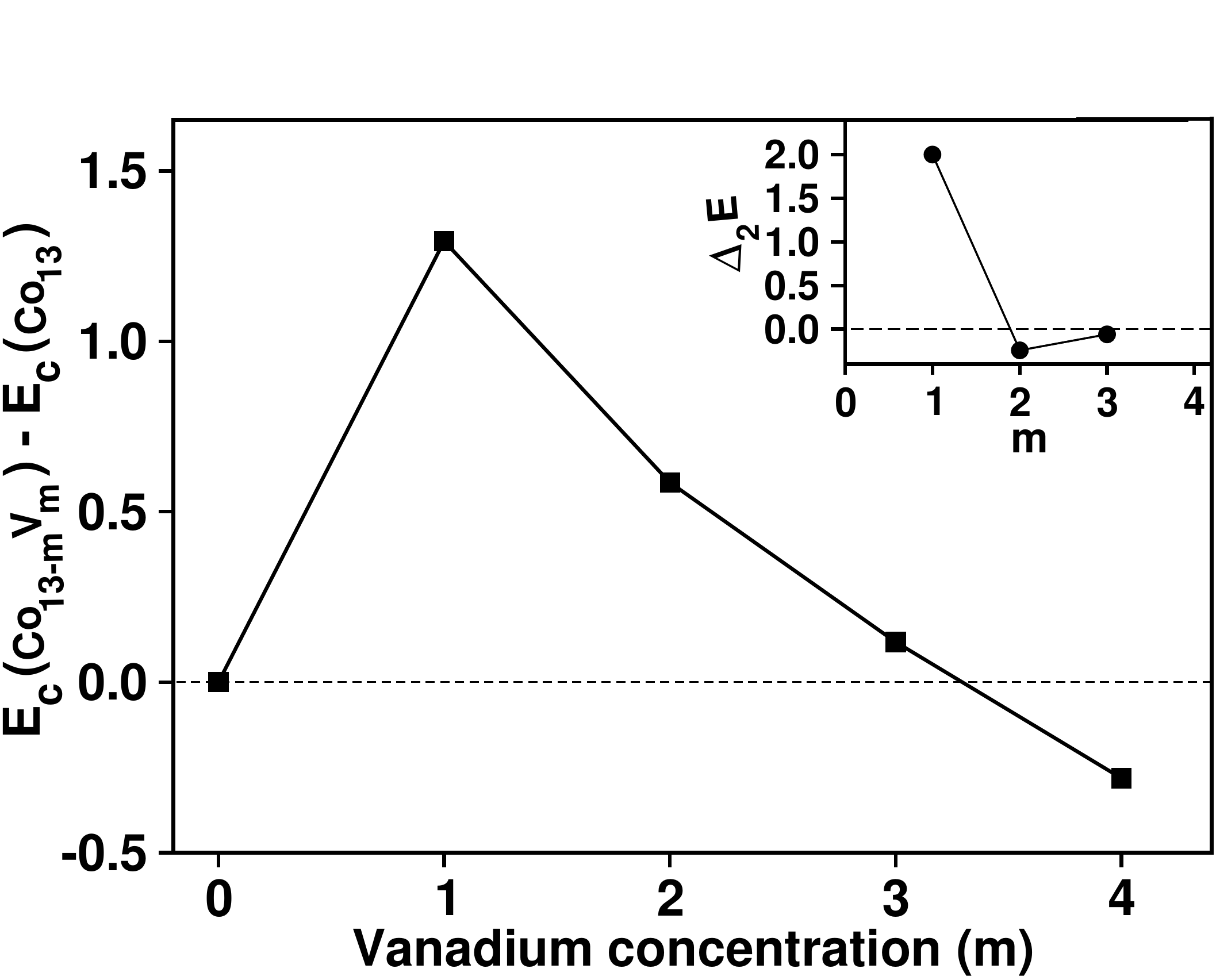}}
\caption{Cohesive energy of the minimum energy structures of Co$_{13-m}$V$_m$ clusters with respect to
 the cohesive energy of minimum energy pure Co$_{13}$ cluster. The dashed line is the reference fixed at the
cohesive energy of pure Co$_{13}$. Inset shows the second difference ($\Delta_2$E) in
 cohesive energy [as defined in Eqn. (\ref{d2e})] for Co$_{12}$V, Co$_{11}$V$_2$ and Co$_{10}$V$_3$.}
\label{fig:cohesive}
\end{center}
\end{figure}

Fig. \ref{fig:cohesive} shows the plot of the cohesive energies of the
 MES of the Co$_{13-m}$V$_m$ clusters with increasing V concentration. The plot is with respect to the cohesive
 energy of the MES of Co$_{13}$ cluster. It is seen that the binding has increased considerably
 compared to that of the optimal Co$_{13}$ cluster  with single V atom doping. However, the cohesive
 energies decrease sharply for the higher concentration of V-atoms. For the double and triple V-atom doping, the cohesive energies 
are above the dashed line indicating their higher stability compared to the pure Co$_{13}$, while for the fourth
 V-atom doping, the cohesive energy is even lower than that of the pure Co$_{13}$. The relative
stability among the clusters of nearby concentrations is more distinct when we plot the second difference in cohesive energies in the inset of Fig. \ref{fig:cohesive}. We use Eqn. \ref{d2e} to calculate the second difference. A sharp pick in the $\Delta_2E$
 at $m =$ 1 points to the exceptional stability of the single V-doped cluster compared to the undoped or more than one V-doped clusters.

In order to understand the stability of the doped clusters, we have calculated the cohesive energies and the bond lengths of the Co-Co, Co-V and V-V dimers as shown in Table \ref{table6.1}. It is seen that the V$_2$ dimer is the most stable and the bond length of the V$_2$
 dimer is also about 14\% shorter than that of the Co$_2$ dimer which has the least binding among the three. On the other hand, the cohesive
 energy and bond length of the Co-V dimer are intermediate of the Co$_2$ and the V$_2$ dimers. For the Co$_{12}$V cluster, we have seen that the center doping
 in icosahedral structure is the most favorable. Binary clusters are known to have the problem of segregation, where the doped atom
 can segregate at the surface or center. To a first approximation, if one of the homonuclear bonds is the strongest, then that species
 tends to be at the center of the cluster.\cite{review1} Our finding of central V-doping is in accordance with this, as the V$_2$ dimer 
has the strongest binding. The atomic radius of a Co-atom and a V-atom, being almost same, the substitution of the central Co atom in an icosahedral Co$_{13}$ cluster by a V atom leaves the structure almost unperturbed. As for evidence, the center to vertex average
 distance and the average distance between two nearby vertices in the optimal icosahedral structure of the Co$_{13}$ cluster are
 2.334 {\AA} and 2.450 {\AA}. For the optimal Co$_{12}$V cluster, these values are almost same, 2.344 and 2.465 {\AA} respectively.
 Therefore the large gain in cohesive energy of the Co$_{12}$V cluster over that of the Co$_{13}$ cluster may be due to the
 enhanced cohesive energy of the CoV dimer over that of the Co$_2$ dimer, though an atom in a dimer is quite different from an atom at the center of a 13-atom cluster. The V atom being at the central position, there are maximum
 number of CoV dimers in V-doped clusters.

\begin{table}[t]

\begin{center}
\caption{\label{table6.1}Cohesive energies and bond lengths of Co$_2$, Co-V and V$_2$ dimers in the present calculation.
 For comparison, we have also listed the experimental values for Co$_2$ dimer (Ref. 35
) and V$_2$ dimer (Ref. 36
).}
\vskip 0.2cm
\begin{tabular}{ccccccc}
\hline
\hline
Dimer    & & \multicolumn{2}{c} {Cohesive energy (eV/atom)}& & \multicolumn{2}{c}{Bond length (\AA)} \\
\cline{3-4}\cline{6-7}
         & &       Theory     &     Expt.                 & & Theory        &   Expt.                \\
\hline
Co$_2$    & &     1.45         &  1.72                     & &  1.96         &  2.31       \\
CoV       & &      1.53        & $\dots$                   &  & 1.87         &  $\dots$ \\
V$_2$      & &     1.81         &  2.47                     & &  1.72         &  1.77  \\
\hline
\hline
\end{tabular}
\end{center}
\end{table}

For the clusters doped with more than one V atom, the favorable structure is again found to be of an icosahedral motif, in which the
 central site is always possessed by a V-atom and the other V-atoms lie on the surface. The number of the surface V atoms and the V-V bonds, therefore, increases with increasing V doping. It favors better binding. On the other hand, the presence of the V-atoms
 on the cluster surface distorts the cluster geometry and increases the center to vertex average distance as well as the vertex-vertex average distance. Effectively, the cluster volume increases with the increasing V-concentration, which destabilizes the
 structure. In Table \ref{table6.2}, we have listed the average distances between center to vertex as well as between two
 nearby vertices for the MES of Co$_{13-m}$V$_m$ ($m =$ 0-4). We believe that the distortion in the structures is due to the charge
 density variation induced by the presence of the V-atoms on the cluster surface. The strained
 cluster structures resulted from distortion can be realized from some open bonds in Fig. \ref{strCo10V3} and Fig. \ref{strCo9V4}. 
Therefore, though the number of the V-V bonds increases, which has better binding compared to the Co-Co or Co-V bonds, the overall cohesive energy decreases monotonically as we go from Co$_{12}$V to Co$_{9}$V$_4$
 by successive doping of V-atoms. It is then obvious that the cluster geometry and the distribution of atoms on the cluster
 surface play an important role in deciding the cluster stability.

\begin{table}
\begin{center}
\caption{\label{table6.2}The average distances in {\AA} between center to vertex and between two nearby vertices
for the minimum energy structures of all the studied clusters. For Co$_{13}$, the values correspond to
 the optimal icosahedron.}
\vskip 0.2cm
\begin{tabular}{ccccccc}
\hline
\hline
Bonds        & & Co$_{13}$ & Co$_{12}$V & Co$_{11}$V$_2$ & Co$_{10}$V$_3$ & Co$_{9}$V$_4$ \\
\hline
center-vertex && 2.334     &  2.344    &  2.354         &   2.375       &   2.398        \\
vertex-vertex && 2.455     &  2.465    &  2.476         &   2.499       &   2.509       \\
\hline
\hline
\end{tabular}
\end{center}
\end{table}

 At this point, it is interesting to study the structure and energetics of the pure V$_{13}$ cluster. Since experiment
 hints symmetric structure for the V$_{13}$ cluster \cite{v13}, we considered an icosahedral geometry and optimized it 
considering {\it all possible} spin configurations. Interestingly, the decreasing trend of cohesive energies for the Co$_{13-m}$V$_m$ clusters
 with increasing V-concentration continues also at $m$ = 13 {\it i.e} pure V$_{13}$ cluster. The cohesive energy of the optimal
 V$_{13}$ cluster is found to be lower than that of the pure Co$_{13}$ cluster by about 3.73 eV. It indicates that V$_{13}$ would be more reactive than any
 of the V-doped Co$_{13}$ clusters. This tendency may be rationalized by considering the case of bulk V and Co, where the bulk V is more reactive
 than Co because of the lower $d$-band filling, relatively higher position of $d$-band center and larger $d$-band width of V compared to Co. 
However, pure V does not have importance in catalysis because its high reaction tendency with absorbate tends to
 poison the surface, without leaving any active site. The desorption energy is also high because of strong V-absorbate bond, which is
 again not favorable for catalysis.

\subsection{Spin gap}
 Analogous to HOMO-LUMO gap of a nonmagnetic cluster, one can define spin gaps for a magnetic cluster as,
\begin{center}
\begin{eqnarray}
\delta_1 = - \left[\epsilon^{\rm majority}_{\rm HOMO} - \epsilon^{\rm minority}_{\rm LUMO}\right]\nonumber\\
\delta_2 = - \left[\epsilon^{\rm minority}_{\rm HOMO} - \epsilon^{\rm majority}_{\rm LUMO}\right]
\end{eqnarray}
\end{center}
and the system is said to be stable if both $\delta_1$ and $\delta_2$ are positive {\it i.e} the LUMO of
 the majority spin lies above the HOMO of the minority spin and vice versa. These represent the energy required
 to move an infinitesimal amount of charge from the HOMO of one spin channel to the LUMO of the other. So magnitude of
 spin gaps is a measure of chemical activeness of clusters. Higher gap means less reactive and vice versa.
The positions of the HOMO and LUMO in both the spin channels and the values of $\delta_1$ and $\delta_2$ for the
optimized structures of 13-atom V-doped Co clusters of all compositions considered here  are given in Table \ref{table6.3}.
\begin{table}
\begin{center}
\caption{\label{table6.3}Positions of HOMO and LUMO in both the spin channels and the values of $\delta_1$ and $\delta_2$
 for the optimized structures of all compositions.}
\vskip 0.3cm
\begin{tabular}{cccccccccc}
\hline
\hline
Cluster   & & \multicolumn{2}{c} {Majority channel} & &\multicolumn{2}{c} {Minority channel} &  &\multicolumn{2}{c} {Spin gap (eV)} \\
\cline{3-4}\cline{6-7}\cline{9-10}
          &  &    HOMO         &    LUMO          &   &      HOMO        &      LUMO        &  &  $\delta_1$ &  $\delta_2$\\
\hline
Co$_{13}$  & &  -3.60        &    -3.35        &  &      -3.48     &    -3.34         &  &   0.26  &  0.13    \\
Co$_{12}$V & &  -3.93        &    -2.61         & &      -3.39     &    -3.34         &  &   0.59  &  0.78    \\
Co$_{11}$V$_2$ & & -3.71      &    -2.83        &  &      -3.43     &    -3.32         & &    0.39  &  0.60    \\
Co$_{10}$V$_3$ & & -3.41      &    -2.98        &  &      -3.34     &    -3.26         & &    0.15  &  0.36    \\
Co$_{9}$V$_4$  & & -3.46      &    -3.02        &  &      -3.33     &    -3.17         & &    0.29  &  0.31    \\
\hline
\hline
\end{tabular}
\end{center}
\end{table}

It is seen that both the spin gaps $\delta_1$ and $\delta_2$ are positive for all the clusters.
 Also Co$_{12}$V has the maximum value of $\delta$'s which again indicates the highest stability of Co$_{12}$V cluster
 compared to the others. Because of this large spin gap, Co$_{12}$V has very low reaction tendency towards
 H$_2$ molecules. However, gap values decrease with increasing V concentration
 and therefore reactivity also increases as observed experimentally.\cite{nonose}

\begin{figure*}
\begin{center}
\vskip 0.5cm
\rotatebox{0}{\includegraphics[height=12.0cm,keepaspectratio]{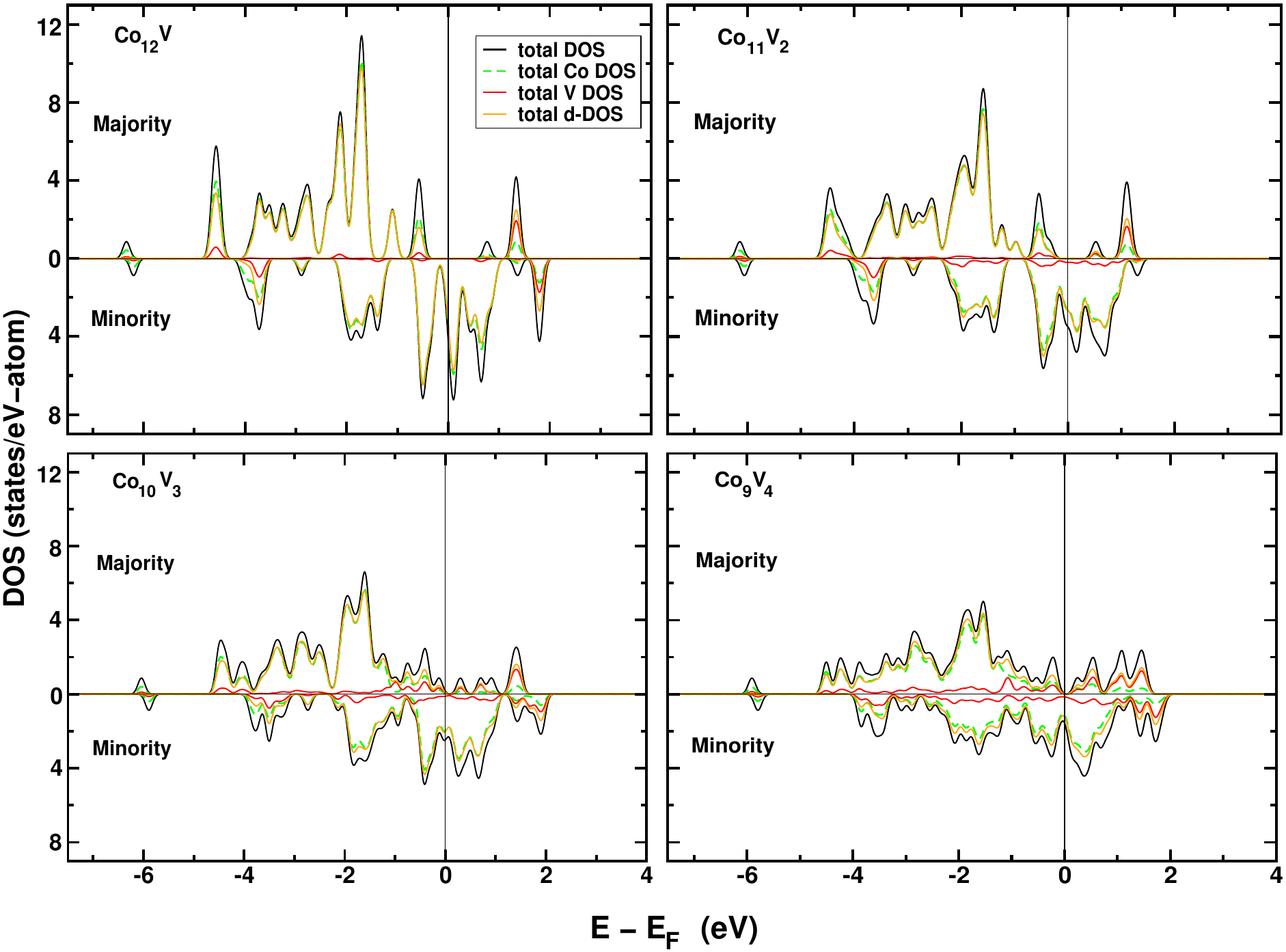}}
\caption{(Color online) Total DOS per atom, Co projected DOS per atom, V projected DOS per atom and
 total d-projected DOS per atom for the optimal structures of Co$_{12}$V, Co$_{11}$V$_2$, Co$_{10}$V$_3$ and
 Co$_{9}$V$_4$. The smearing width is fixed at 0.1 eV. Vertical line through zero is the Fermi energy (E$_F$).}
\label{totaldos}
\end{center}
\end{figure*}

\begin{figure*}[t]
\begin{center}
\vskip 0.9cm
\rotatebox{0}{\includegraphics[height=8.0cm,keepaspectratio]{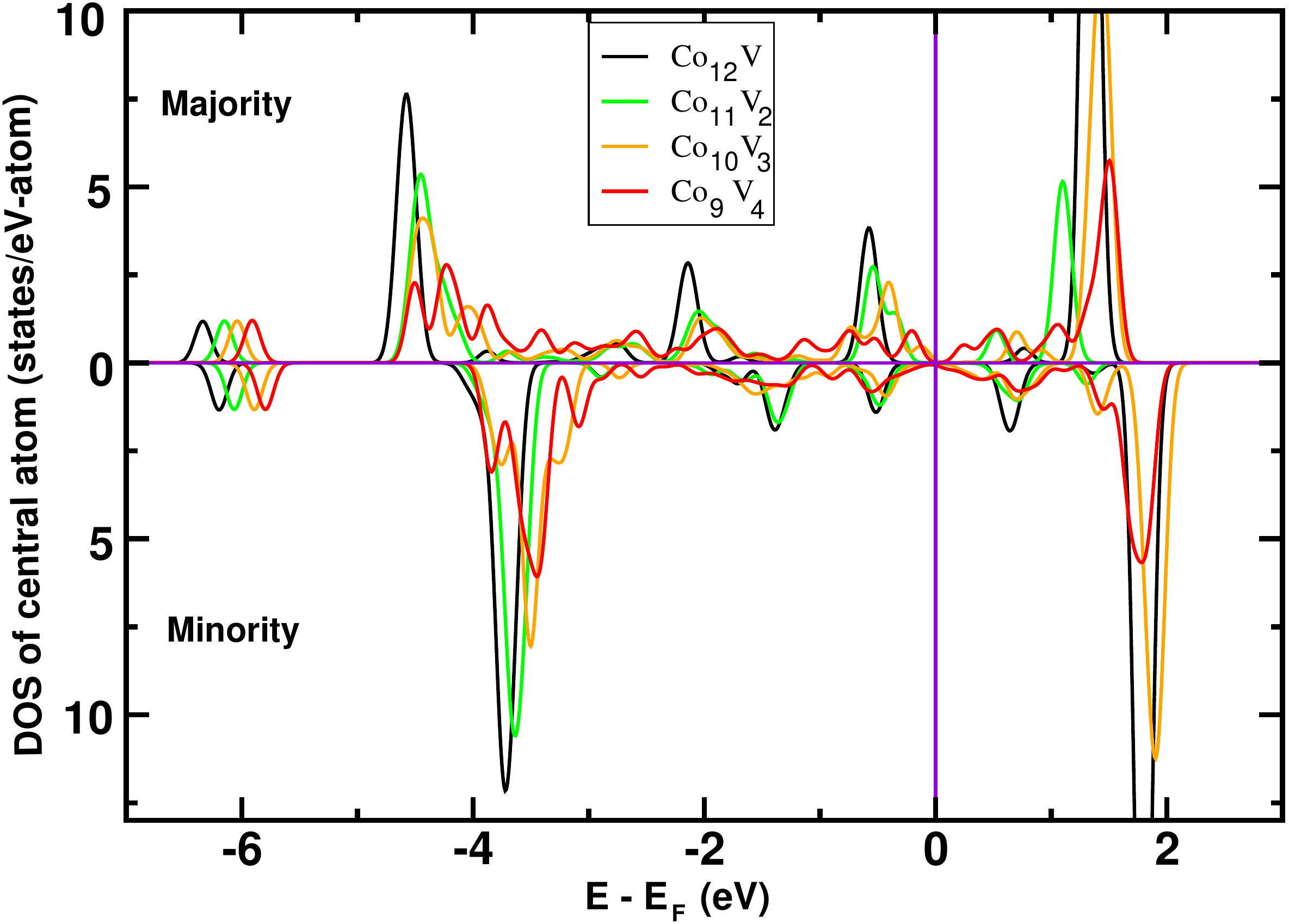}}
\caption{(Color online) Total DOS of the central V atom in the optimal structures of Co$_{12}$V, Co$_{11}$V$_2$, Co$_{10}$V$_3$ and
 Co$_{9}$V$_4$. The Fermi energy (E$_F$) is fixed at zero.}
\label{centreatomdos}
\end{center}
\end{figure*}

\subsection{Density of states (DOS)}
Fig. \ref{totaldos} shows the spin-polarized total DOS, Co projected DOS, V-projected DOS and total $d$-orbital projected DOS for the
 MES of Co$_{12}$V, Co$_{11}$V$_2$, Co$_{10}$V$_3$ and Co$_{9}$V$_4$ clusters. It is seen that the total DOS and the total $d$-projected DOS are almost overlapping each other. This indicates that the cluster properties are mostly dominated
 by the $d$-electrons which is generally expected for transition metal clusters. The trend in structural stability and reactivity
 of the clusters as discussed in the previous sections, is also obvious from the nature of the total DOS. The majority spin channel
 of each cluster has a gap. This gap is maximum for the Co$_{12}$V cluster and it decreases as the number of surface V-atoms increases.
 In the minority spin channel, there is finite amount of states at the Fermi energy and these are contributed solely by the exterior surface atoms,
 as the central atom does not have any contribution at the Fermi energy (cf. Fig. \ref{centreatomdos}). It is also
seen that the states are more localized in case of the Co$_{12}$V cluster, while they are gradually spreading out and therefore the DOS height,
 specially in the majority spin channel, decreases with the increasing V-concentration. At this point it is relevant to mention a common
 feature observed in case of extended surface. The gaseous molecules absorb well on clean surface of an early transition metal as well
 as on the surface with a multilayer of late transition metal, but not on the surface with a monolayer of late transition metal. For example, the photoemission experiments by El-Batanouny {\it et al} \cite{batanouny} showed that a $H$ atom adsorbs well both on the clean Nb(110) surface
 as well as on the surface with a multilayer of Pd, but does not adsorb on Nb(110) surface with a monolayer of Pd and this is due to the
 decrease of density of states of $d$-electrons of Pd near the Fermi level. Similar behavior had been observed also in case of a monolayer
 of Pd on W(110) surface, in the experiment of CO-adsorption.\cite{pdonw}  Although, a cluster of 13 atoms is substantially different in its
 property from that of the surface or bulk, one may resemble the stability of the Co$_{12}$V cluster towards H-adsorption with that of extended surface, where the central $V$ atom of Co$_{12}$V corresponds to
 an early transition metal substrate and the exterior twelve Co atoms correspond to a late transition-metal monolayer.

In order to see the chemical activity of the central V-atom, we have plotted in Fig. \ref{centreatomdos}, the projected total
 DOS of the central V-atom for each of the Co$_{13-m}$V$_m$ clusters. First of all, it is seen that there is no finite
states at the Fermi energy, meaning that the central V-atom is not chemically active. Also each of the
majority and minority spin channels has two large pecks: one is deep below the Fermi energy and another is above the 
Fermi energy. However, the peck heights gradually decrease, the states are broadened out and shifted towards higher
 energy with increasing V-concentrations. It is therefore indicating that the presence of surface V-atoms induces some sort of chemical
 activeness to the central atom.

\section{Chemisorption with H$_2$ molecules}
In order to gain some understanding about the cluster chemical reactivity, we have investigated the
 chemisorbed structures of the Co$_{13-m}$V$_m$ clusters upon H$_2$ adsorbtion. To check the robustness of our chemisorption
 calculations involving H-atoms, we have first calculated the cohesive energy,
 bond length and vibrational frequency of H$_2$ dimer. Our calculated values have been listed in Table \ref{h2dimer}.
 These values are  typical for gradient-corrected calculations of H$_2$, which have been  done
 before \cite{h2dimer_theory} and they agreed reasonably well with experiment.\cite{h2dimer_expt}

\begin{table}
\begin{center}
\caption{\label{h2dimer}Theoretical values of cohesive energy, bond length and vibrational frequency for H$_2$ dimer in the present calculation. Experimental values in Ref. 41 are also given for comparison.
}
\vskip 0.3cm
\begin{tabular}{cccc}
\hline
\hline
        &  Cohesive   &  Bond length ({\AA}) &  Vibrational \\
        & energy (eV) &                      &  frequency (cm$^{-1}$) \\
\hline
Theory  &  4.520                 &  0.752               &         4339   \\
Experiment &  4.750              &   0.741              &         4395    \\
\hline
\hline
\end{tabular}
\end{center}
\end{table}

\begin{figure*}[t]
\begin{center}
\vskip 0.6cm
\rotatebox{0}{\includegraphics[height=3.5cm,keepaspectratio]{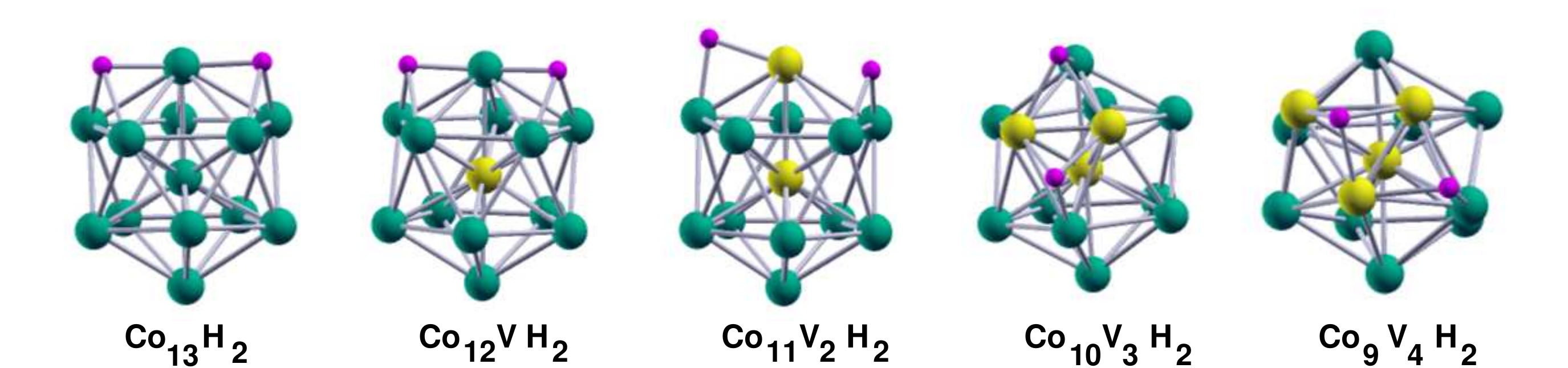}}
\caption{(Color online) The calculated minimum energy chemisorption structures with H$_2$ on the minimum energy structures
 of Co$_{13-m}$V$_m$ ($m =$ 0-4). It is clearly seen that hollow site on the surface is preferred by
chemisorbed hydrogen. Color convention is same as in Fig. \ref{fig:strCo12V} with pink colored dots representing H atom. }
\label{chemisorp}
\end{center}
\end{figure*}

We have then performed an exhaustive search for the MES, taking H$_2$ at different possible places
on the MES of the corresponding bare cluster for each composition. Fig. \ref{chemisorp} shows our
calculated lowest energy chemisorption structures after H$_2$ absorbsion for clusters of all compositions. First thing to notice is that
H$_2$ molecule chemisorbs dissociatively in each case, {\it i.e} the distance between the two H-atoms in the chemisorption
 structures is much larger than the H-H bond length of an isolated H$_2$ molecule. The chemisorption
 gives rise to moderate perturbation to the structures and makes them more symmetric ({\it i.e} surface bonds are now
less dispersive) compared to the parent structure without hydrogen. There are three
possible ways for the H-atoms to be adsorbed on each cluster : on top of an atom (one fold), at a bridge position between two
 atoms (two fold) and at the hollow site of a triangular plane on the cluster (three fold). Again, as the surface contains two
 species of atoms for Co$_{11}$V$_2$, Co$_{10}$V$_3$ and Co$_{9}$V$_4$, then the one fold position can be on top of a
 surface V-atom or on top of a surface Co-atom. Two fold position can be the top of a V-V bond, Co-V bond or a Co-Co bond which
 can be nearer to or away from the V-site. Similarly, in triangular plane, there are several possibilities : ($i$) all
 the three atoms can be V-atoms (only possible for Co$_{9}$V$_4$), ($ii$) one Co atom and two V-atoms, ($iii$) two
 Co-atoms and one V-atom, and ($iv$) all three are Co-atoms. We have considered all these possible combinations during optimization.
 It is, however, seen that in each of the optimized structures, H-atoms absorb at the hollow site on the surface and
 for m $\ge$ 2, they prefer the association of the local V atom. For example in the Co$_{9}$V$_4$H$_2$ cluster, one H-atom absorbs
 at the hollow site of V-V-V triangular plane and the other H atom on top of a V-V-Co triangular plane. For Co$_{11}$V$_2$H$_2$, the
 H-atoms appear to absorb at the bridge positions, but they are inclined with an angle such that the absorption tends
toward a three fold configuration. The preference of more coordinated hollow site is likely due to geometric arrangements.
 It allows the H-atom to interact more with the V or Co atoms. On the other hand, the V-site preference of H-atom is due to the formation of stronger $s$-$d$ bond with V-atom compared to that with Co-atom.

The higher reactivity of the Co$_{13}$ cluster compared
 to the Co$_{12}$V cluster, while both of them have the same twelve Co-atoms on the surface, can be understood from charge transfer analysis.
We have computed the charge enclosed within a sphere around each surface atom of the optimal structures of Co$_{13}$H$_2$ and
 Co$_{12}$VH$_2$ clusters. It is seen that the amount of charge on the surface Co-atoms in the Co$_{13}$H$_2$ cluster is larger by about 0.2 (in unit of 
electronic charge) than
 that of Co$_{12}$VH$_2$ and it is mainly contributed by the $d$-electrons of surface Co atoms. Therefore stronger 3$d$-1$s$ interaction between
 surface Co-atoms and chemisorbed H-atoms in case of Co$_{13}$H$_2$ increases its reactivity. This result is in accordance with
 that of Fujima {\it et al} \cite{fujima} who also predicted the stronger interaction between 1$s$ of H atom and the 3$d$ orbital
 of the surface Co atom in the Co$_{13}$H$_2$ cluster compared to that in the Co$_{12}$VH$_2$ cluster using density of state analysis. They showed
 that the anti-bonding orbital component between the H-1$s$ electron and the 3$d$ electron of the surface Co-atoms shifts further away above the 
HOMO in case of Co$_{13}$H$_2$ compared to that of Co$_{12}$VH$_2$, while the bonding orbital components for both the clusters stay almost at the same energy position below the HOMO. Extending the charge transfer analysis for the Co$_{13-m}$V$_m$H$_2$ clusters with $m$ = 2,3 and 4, we find that the average charge per surface V-atom is gradually increasing with the increase of V-concentration which indicates the formation of stronger V-H bonds at the surface with increasing V-doping. Average charge per surface Co-atom for these three
 clusters, however, remains almost the same.

\begin{figure}
\begin{center}
\vskip 0.9cm
\rotatebox{0}{\includegraphics[height=5.5cm,keepaspectratio]{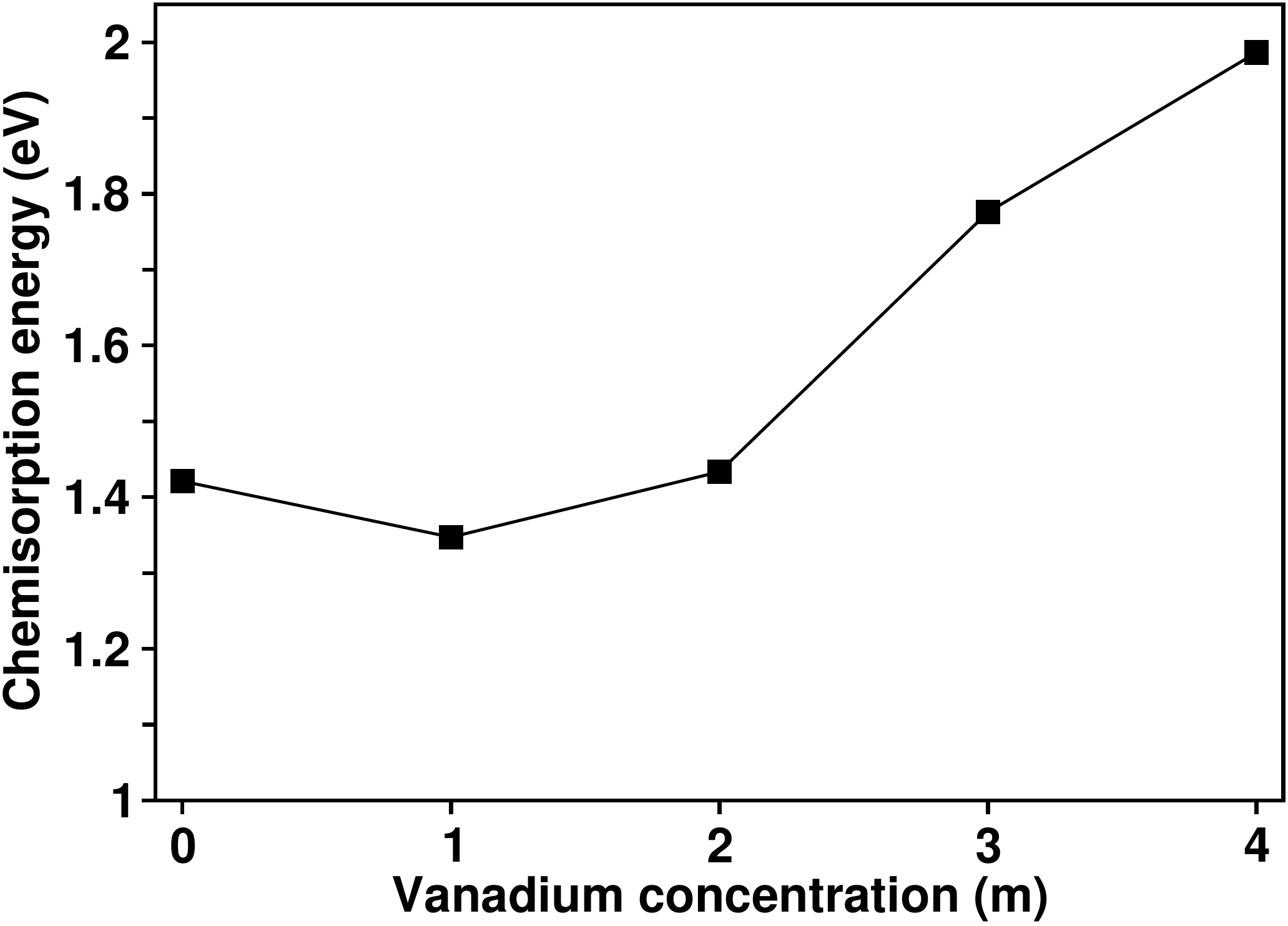}}
\caption{The calculated chemisorption energy (using Eqn. \ref{chem_energy}) for Co$_{13-m}$V$_m$H$_2$ clusters.}
\label{chemisorption}
\end{center}
\end{figure}

Fig. \ref{chemisorption} shows the plot of chemisorption energy with increasing V-concentration, where the
 chemisorption energy is defined as
\begin{equation}
 D_e(E) =  E({\rm Co}_{13-m}{\rm V}_m) + E({\rm H}_2) - E({\rm Co}_{13-m}{\rm V}_m{\rm H}_2)
\label{chem_energy}
\end{equation}
It is a positive quantity and gives a measure of binding strength of H$_2$ molecule with the cluster. The plot shows a minimum for Co$_{12}$VH$_2$ 
 again indicating the lowest binding efficiency of Co$_{12}$V with H. However, chemisorption energy
 increases with increasing V concentration. The source of this chemisorption energy is the cluster
 rearrangement energy ({\it i.e} the energy change due to the geometrical rearrangement of the cluster upon chemisorption)
 and the efficient cluster-absorbate bonding in presence of V due to more efficient $s$-$d$ hybridization.

\section{\label{summary}Summary and Conclusions}
To summarize, we have studied the geometric and electronic structures
 of V doped Co$_{13}$ clusters and their chemisorption towards hydrogen molecules using first
 principles density functional calculation. The lowest energy structures of all compositions prefer to have
icosahedral geometry, unlike hexagonal symmetry preference of the pure Co clusters. For the Co$_{12}$V cluster, the single
V atom prefers to sit at the central site, thereby guarded by all the surface Co atoms and cannot participate directly
 in the chemisorption reaction. Consequently reactivity of Co$_{12}$V becomes very less. On the other hand, for more
 than one V atom doping, the additional V-atoms reside on the surface and come in direct contact with chemisorbed
 H-atom and reactivity increases. Our calculated spin gaps, density of states and charge transfer analysis explain nicely
 the stability of clusters and their tendency towards chemisorption. In the chemisorbed structures, H-atoms adsorb dissociatively at
 the more coordinated hollow sites and they prefer V-site association due to stronger 3$d$-1$s$ hybridization. To have better insight into the chemisorption reaction, one needs to study the transition states for the optimal cluster of each composition and we believe
 that calculation of activation barriers will also lead
 to same conclusion as we have predicted here.

\acknowledgments
 S. D. is thankful to Council of Scientific and Industrial Research (Government of India) for financial support.
 T.S.D. thanks Department of Science and Technology, India for Swarnajayanti fellowship and the support through
Advanced Materials Research Unit.

\end{document}